\documentclass[prl,preprint,amssymb,amsfonts,superscriptaddress]{revtex4}
\usepackage{epsfig}
\usepackage{amsmath}

\begin{document}

\title{Artificial trapping of a stable high-density dipolar exciton fluid}

\author{Gang Chen, Ronen Rapaport, L. N. Pffeifer, K. West, P. M. Platzman, Steven Simon}
\affiliation{Bell Laboratories, Lucent Technologies, 600 Mountain
Avenue, Murray Hill, New Jersey 07974}
\author{Z. V\"or\"os, and D. Snoke}
\affiliation{Department of Physics and Astronomy, University of
Pittsburgh, Pittsburgh, Pennsylvania 15260, USA}

\begin{abstract}

%Exciton fluids in semiconductors are predicted to undergo quantum
%phase transitions to a superfluid or Bose-Einstein condensed state
%at low temperatures and high densities.

We present compelling experimental evidence for a successful
electrostatic trapping of two-dimensional dipolar excitons that
results in stable formation of a well confined, high-density and
spatially uniform dipolar exciton fluid. We show that, for at
least half a microsecond, the exciton fluid sustains a density
higher than the critical density for degeneracy if the exciton
fluid temperature reaches the lattice temperature within that
time. This method should allow for the study of strongly
interacting bosons in two dimensions at low temperatures, and
possibly lead towards the observation of quantum phase transitions
of 2D interacting excitons, such as superfluidity and
crystallization.

\end{abstract}

\maketitle

%\textbf{ Two-dimensional dipolar excitons, i.e., Coulomb-bound but
%spatially separated electron-hole pairs, are new promising
%candidates for the realization of superfluidity in semiconductors
%\cite{KosterlitzJPhysC1973,LozovikJETP1975,TejedorPRL1997,KeldyshJETP1968},
%as their extremely long lifetime should make their cooling and
%thermalization feasible. However, a rapid dipolar exciton
%expansion \cite{VorosPRL2005,RapaportPRB2006} can be a major
%obstacle in realizing such quantum phases since it prevents the
%formation of a stable and high-density exciton fluid. Here, we
%realize a scheme for trapping excitons electrostatically under a
%local gate \cite{RapaportPRB2005}. We report on compelling
%evidence that the trapped exciton fluid is long-lived, highly
%stable, spatially uniform, and can reach densities much higher
%than those achieved without a trap. This ability to trap excitons
%in a well-controlled and flexible electronic device should allow
%for more general studies of various possible phases of strongly
%interacting bosons in two dimensions.}

The different thermodynamic phases of a system of particles and
the transitions between them is fundamental to our understanding
of the material world. Most notably, the dimensionality of the
particle system, together with the importance of quantum
statistical effects, can have a profound impact on its physical
behavior. Two-dimensional (2D) systems of interacting particles
are expected to have a rich and unique behavior. They are,
however, experimentally much more challenging to realize than
their three dimensional counterparts, and 2D bosons, especially in
solid state systems, are essentially unexplored experimentally.

One of the most promising 2D system of bosons is that of 2D
excitons, which are coulomb bound electron-hole pairs confined in a
semiconductor quantum well layers. In the dilute limit, where
$n_X^{-1/2}\gg a_X$ ($n_X$ being the exciton 2D density and $a_X$ is
the Bohr radius), excitons can be considered as bosons, as the
identity of their fermionic constituents is essentially hidden
within the excitonic atomic-like structure. It has been predicted
that 2D excitons at low temperatures can undergo a quantum phase
transition to either a Bose-Einstein condensate
\cite{KeldyshJETP1968} similar to trapped alkali atoms
\cite{KetterleNature2002}, or to a condensed superfluid state
\cite{KosterlitzJPhysC1973,LozovikJETP1975,TejedorPRL1997}, similar
to, for instance, superfluid $^4$He. A clear observation of this
elusive phase remains one of the tough but most interesting
challenge that low-dimensional systems physicists have been working
toward for the past several decades
\cite{ExcitonReviews,EisensteinNature2004}.

During the past decade, a unique class of 2D excitons, that are
spatially indirect, has been introduced to this field of study. In
these spatially indirect systems, the constituent electrons and
holes are separated into two weakly coupled quantum wells with an
inter-well energy barrier
\cite{FukuzawaPRL1990,AlexandrouPRB1990}. This separation results
in two unique features. First, the exciton lifetime can become
extremely long (in the microseconds regime), a crucial requirement
for thermalization processes
 \cite{RapaportPRL2004,ButovPRL2004,RapaportPRB2006}. Secondly, the
indirect excitons are intrinsically dipolar, and their dipole
moments are \emph{all} aligned perpendicular to the quantum wells
plane, resulting in a net dipole-dipole repulsive interaction
between them. This short range interaction is expected to play a
crucial role in the physics of quantum phase transitions of such an
exciton fluid, which is a good model system for interacting bosons
in two dimensions. However, as we have recently suggested, the
exciton permanent dipole moment also prevents the optically excited
dipolar excitons from forming a stable fluid and from maintaining
densities high enough for a quantum phase transition to occur
\cite{RapaportPRB2006}, as this exciton fluid tends to rapidly
expand under its own repulsive forces
\cite{VorosPRL2005,RapaportPRB2006}. This expansion can prevent
other possible interesting phases such as exciton crystallization
\cite{SchmitPRL2002}, which requires a stable and well confined
fluid.

It is tempting to try and exploit several of the concepts that
were so successful in atomic physics in order to overcome the
expansion problem and achieve a high-density, stable and
long-lived exciton gas. We have recently proposed a scheme for
trapping a high-density exciton fluid in a quantum well plane
\cite{RapaportPRB2005,oneDtrap}. This scheme utilizes the
interaction of the exciton dipoles with an externally applied
local electric field to create an effective barrier that prevents
the exciton fluid from expanding. In this report, we present
compelling evidence for the successful implementation of our
scheme, and show that in the presence of such a dipolar exciton
trap, a high-density, stable, and spatially uniform exciton fluid
is maintained in a confined and controlled configuration. We show
that, for at least half a microsecond, the exciton fluid sustains
a density higher than the critical density for degeneracy if the
exciton fluid temperature reaches the lattice temperature within
that time. We believe this demonstration is a major step towards
the observation of quantum phase transitions of 2D interacting
excitons and the possible exciton crystallization.

\begin{figure}[htb]
\begin{center}
\includegraphics[scale=0.4]{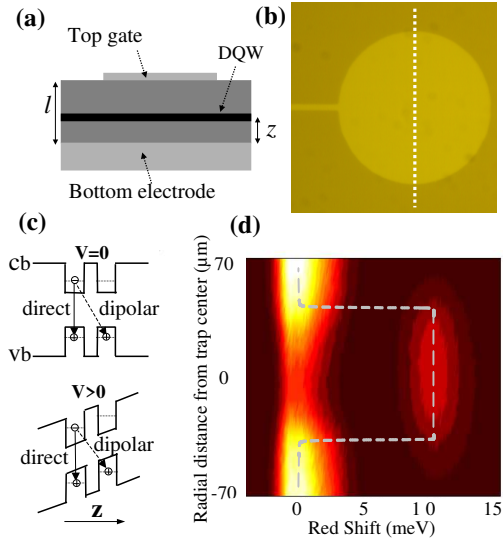}
\caption{(Color online)(a) A schematic of the Xtrap configuration.
(b) A microscope image of an Xtrap upper gate with a diameter
$D=50\mu$m. The wire is $2\mu$m wide. (c) The DQW band diagram
under the top gate in the absence (upper) and the presence (lower)
of an applied gate bias. Outside the gate, the band is flat,
regardless of the bias. The two arrows represent the direct and
indirect exciton transitions. (d) Spatially resolved PL spectra
from an $80\mu$m Xtrap (sample A) collected from a radial
cross-section of the top gate (along the white dotted line in
(b)), with a defocused He-Ne excitation and a bias of 3V. The
dashed white line is the calculated electrostatic potential. The
energy axis shows the red shift of the emission from the direct
exciton emission line whose actual position is at 1.5567 eV (797
nm).}\label{figure1}
\end{center}
\end{figure}

We follow our exciton trap (Xtrap) concept and design developed in
Ref.~\onlinecite{RapaportPRB2005}, where dipolar excitons are
trapped under local electrostatic gates. GaAs/AlGaAs double
quantum well (DQW) structures are grown on top of an n$^+$ doped
GaAs substrate \cite{sample_details}. Circular semi-transparent Ti
gates (20nm thick, see Fig.~\ref{figure1}b) with different
diameters ($D=4-80\mu$m) are deposited on top of the DQW samples.
An electric bias (typically few volts) is applied to each of the
gates via a narrow Ti evaporated wire (1-2 $\mu$m wide). The
conductive substrate is used as the bottom electrode. This
configuration is illustrated in Fig.~\ref{figure1}a. The sample
thickness between the two electrodes is $l$, while the DQW
structure is grown at a distance $z$ as measured from the bottom
electrode. Two samples are described here. Both have a similar DQW
structure and similar gate designs, and the main difference
between them is their $z/l$ ratio. One (sample A) has $z/l\simeq
0.03$ while the other (sample B) has $z/l\simeq 0.3$.

\vspace*{0cm}

The dipolar exciton trapping takes place due to their interaction
with the applied electric field under the biased Xtrap gate:
dipolar excitons (with a dipole moment $\vec{d}_{X}=-ez_0\hat{z}$
\cite{sample_details}), are "high field seekers", i.e., when they
are aligned parallel to an electric field $\vec{E}$, they will
always seek its maximum, as their total energy is reduced by the
dipole-field interaction term: $\varepsilon_{df} = \vec{d}_{X}
\cdot \vec{E}(r_{\|},z)=d_XE_z(r_{\|},z)$. Thus applying an
external bias locally to the gate results in an effective
confining potential for the dipolar excitons under the gate
\cite{RapaportPRB2005}. An alternative way of understanding the
trapping mechanism is to note that, as is shown in
Fig.~\ref{figure1}c, the applied electric field under the gate
tilts the band structure in the $\hat{z}$ direction, hence
lowering the dipolar exciton transition energy locally. To
experimentally measure the Xtrap energy profile, we spatially
resolve the photoluminescence (PL) emission energy of the dipolar
excitons, by collecting emission spectra along a radial
cross-section of the Xtrap. An example of such spectral image is
shown in Fig.~\ref{figure1}d, taken from a 80$\mu$m diameter Xtrap
of sample A at T=5K with an applied bias of $3V$, excited using a
defocused cw He-Ne laser at 632.8nm and collected with a liquid
nitrogen cooled CCD camera mounted at the exit port of a
spectrometer. The confining potential profile can be clearly seen
from the red-shifted PL under the gate. The emission energy
outside the trap corresponds to zero field, or "direct" exciton
line at 1.5567eV. The white dashed line is the calculated Xtrap
potential (see Ref.~\cite{RapaportPRB2005}), which nicely fits the
experimental data.

Ideally, for diameter $D\gg l$, such an Xtrap confines a circular
"pool" of dipolar excitons which are free-moving under the trap
gate and experience a strong reflecting force near its boundary.
While this seems simple enough, we have shown before that any
in-plane fields at the Xtrap boundary, arising from the above gate
geometry, can cause rapid ionization of excitons through
field-induced tunneling, thus significantly degrading the Xtrap
quality and making it very leaky \cite{RapaportPRB2005}. Such a
loss of excitons can be represented by an effective Xtrap
lifetime, that can be orders of magnitude shorter than the
intrinsic lifetime of the dipolar excitons, making the trapping
very inefficient. A possible suggested solution
\cite{RapaportPRB2005,ButovCondmat2005}, is to design an Xtrap
with the DQW structure grown very close to the conductive
substrate (corresponding to a small $z/l$ ratio), where the
in-plane components of the electric field rapidly vanish. Since
the ionization rate depends exponentially on the magnitude of the
in-plane field component, a small $z/l$ is expected to be a very
important requirement for a high quality Xtrap, which is crucial
for achieving a high-density, spatially uniform, and long-lived
exciton fluid. In what follows we will present evidence that
indeed, with the right design, such a high quality Xtrap is
feasible.

\vspace*{0cm}
\begin{figure}[htb]
\begin{center}
\includegraphics[scale=0.5]{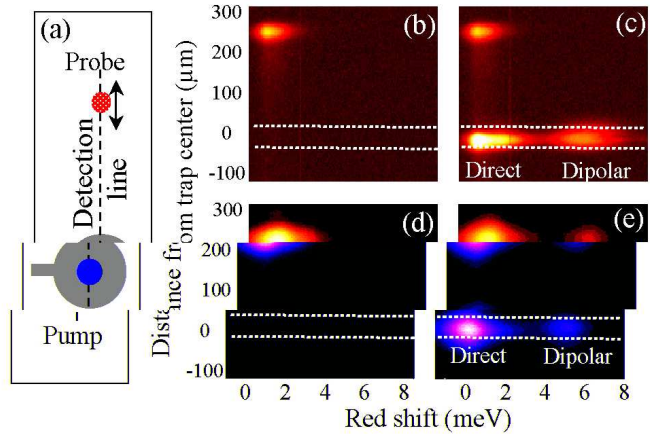}
\caption{(Color online)(a) Geometry of the PL pump-probe
experiment on an $50\mu m$ diameter Xtrap. The PL is spatially
resolved along the dashed line with its energy shown as a red
shift from the direct exciton line. PL spectral images from sample
A with (b) pump off, and (c) pump on. The pump power is 30$\mu W$
and dipolar exciton red shift is 5 meV. PL spectra for sample B
with (d) pump off, and (e) pump on taken under identical
conditions. The white dotted lines mark the trap
boundary.}\label{figure2}
\end{center}
\end{figure}

To investigate the effect of the exciton ionization at the edge of
the trap, we developed a spatially resolved "excitation
pump-probe" technique: an "excitation-pump" from a non-resonant,
cw He-Ne laser ($\lambda=632.8$nm), is focused to the center of an
Xtrap ($r_{\|}=0$). An additional, similar but weak
"excitation-probe" beam is focused outside the trap where no
external bias was applied. The experimental geometry is shown in
Fig.~\ref{figure2}a. The PL of excitons created locally by this
second beam was monitored, and the details of this local PL
spectrum are directly related to the local environment of the
probing excitons. If there is a significant ionization of dipolar
excitons that are created by the excitation-pump within the Xtrap,
the probing excitons created by the weak excitation outside the
trap will interact with the ionized carriers leaking from the
Xtrap and their emission will be spectrally modified.
Fig.~\ref{figure2} presents the spatially resolved PL spectra
along a line going through an $50\mu m$ diameter Xtrap center for
the two samples A and B. The PL energy is measured as a red shift
from the direct exciton transition energy. Fig.~\ref{figure2}b-c
show the such PL spectra from sample A ($z/l=0.03$) with the
excitation-pump (inside the Xtrap) off and on respectively. For
the \emph{entire} range of pump excitation powers and gate bias,
we find that the probe PL is independent of the presence of the
excitation-pump at the center of the trap, even for
excitation-probe positions that are very close to the trap
boundary (not shown). The probe excitons PL always corresponds to
the transition energy and lineshape of direct excitons, consistent
with the zero bias conditions expected outside of the Xtrap and is
an indication of the high quality of the trap boundaries in sample
A. On the contrary, the probe PL behavior of sample B ($z/l=0.3$)
is remarkably different: With pump-excitation off, the PL of the
probe excitons is similar to that observed is sample A, as is
shown in Fig.~\ref{figure2}d. However, as the pump-excitation beam
is turned on, these PL spectra are dramatically modified, and we
observe an additional probe PL peak at the lower energy side of
the direct exciton emission line, as is shown in
Fig.~\ref{figure2}e. This dramatic change of the remote probe PL
can only be explained by a \emph{leakage of carriers} from the
Xtrap to the location of the probe, changing the local
electrostatic environment and thus affecting the emission.

\vspace*{0cm}
\begin{figure}[htb]
\begin{center}
\includegraphics[scale=0.45]{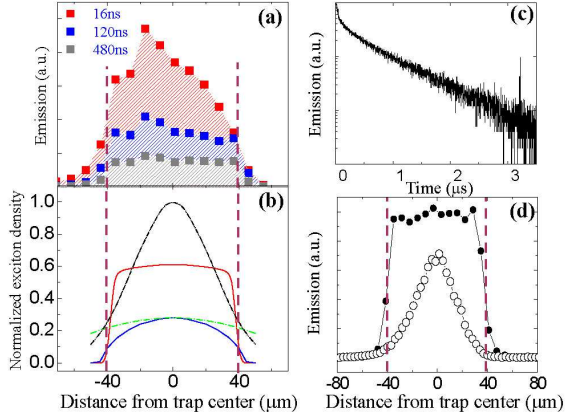}
\caption{(Color online)(a) Spectrally integrated intensity of the
dipolar exciton PL along a radial cross-section of an $80\mu$m
Xtrap of sample A with a potential depth of 30meV, for three
different times after the initial excitation (FWHM$\simeq~30\mu
m$). (b) Calculated exciton density 120ns after excitation for the
case of reflecting Xtrap boundaries (solid-red), absorbing Strap
boundaries (solid-blue), and no boundaries (dash-dot green). The
black dash-dotted curve presents the initial exciton distribution.
(c) The time trace of the spectrally integrated PL at the center
of the trap in (a). (d) Time integrated spatial profile of dipolar
excitons from an 80 $\mu m$ trap of sample A (black dots) and B
(open circles) excited by 10 ns pulses with FWHM$\simeq 40\mu m$.
The dashed lines mark the position of the trap.}\label{figure3}
\end{center}
\end{figure}

To understand the dynamics of the trapped excitons and verify the
trap quality we measured the time dependence of the spatial
cross-sectional profile of the PL from an 80$\mu$m diameter Xtrap
($6V$, $\sim$30meV deep) of sample A, after a short pulse
excitation (pulse width 2ps, repetition rate 250 KHz, spatial
FWHM$\simeq 40\mu$m, and excitation intensity 200 $\mu$W),
resonant with the direct exciton transition. The dipolar exciton
density immediately after the laser excitation pulse is estimated
to be $\sim 8\times 10^{10}cm^2$ (based on a method to be
discussed later). The PL was spectrally integrated over the whole
dipolar exciton spectral line and detected using a single photon
counting photomultiplier tube. The details of this measurement
technique is given in Ref.~\onlinecite{VorosPRL2005}. PL profiles
at several times after the optical excitation are shown in
Fig.~\ref{figure3}a. The boundary of the trap is marked by the
dashed lines. Shortly after the excitation, the PL profile is
nearly Gaussian, already broader but similar in shape to that of
the excitation beam. The PL profile continues to rapidly expand
toward the trap edges and to flatten. After $\sim$100ns the PL
profile becomes completely flat with a sharp drop at the trap
boundaries, only limited by the imaging resolution, and decays
uniformly with time (the small and sharp peak at $\sim$20$\mu$m is
due to scattering of residual laser pulses). A time trace of the
PL at the center of the trap is plotted in the
Fig.~\ref{figure3}c, showing the exponential decay with a
microsecond time constant after the profile has flattened (the
initial faster decay is mainly due to the expansion of the exciton
fluid within the trap). Such a behavior is indeed expected from a
combination of an outward expansion driven by the effective net
dipolar repulsion force and the sharp profile of reflecting forces
\emph{with negligible losses} at the Xtrap boundary.
%Since the
%exciton expansion within the trap occurs on a time scale much
%shorter than the exciton lifetime, a flat dipolar exciton profile
%with sharp boundaries should be observed even in time-integrated
%measurements. Th
The above measurements are consistent with a calculation of the
expansion dynamics in the Xtrap, based on the model we developed
in Ref.~\cite{RapaportPRB2006} for free expansion of dipolar
excitons, but with the addition of an external electrostatic trap
potential barrier\cite{RapaportUnpublished}. The calculated
exciton profile 120 ns after the excitation, shown by the red
curve in Fig.~\ref{figure3}b, nicely reproduces the expansion and
flattening of the exciton distribution. In contrast, the
calculated exciton profile for absorbing trap boundaries (as is
expected if a fast exciton ionization takes place) and that of
expanding excitons with no boundaries, shown by the blue and green
curves respectively, always keep their curvature and do not show
flattening at any stage of the expansion. This is verified by
Fig.~\ref{figure3}d which shows a comparison between the time
integrated spatial emission profile from an 80 $\mu m$ trap of
sample A (black dots) and B (open circles). In addition, the fact
that the decay time of the excitons is similar to their intrinsic
lifetime is a strong evidence that exciton loss at the Xtrap
boundary is insignificant. All of the above findings therefore
confirm high quality Xtraps in sample A, with good reflecting
boundaries and negligible ionization.

For a lattice temperature of 1.4K of the experiment in
Fig.~\ref{figure3}, the critical density for degeneracy is $\simeq
2\times 10^{10}cm^{-2}$. Fig.~\ref{figure3}c shows that for at
least half a microsecond the exciton fluid sustains a density
higher than that critical density.  The exciton fluid is highly
degenerate if it is thermalized with the lattice within that time.

\begin{figure}[htb]
\begin{center}
\includegraphics[scale=0.4]{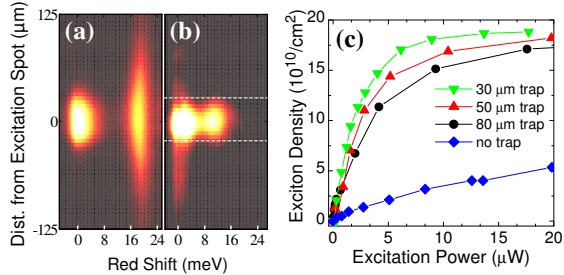}
\caption{(Color online)(a) Spatially resolved PL spectra of (a)
free expanding dipolar excitons, (b) dipolar excitons confined in
a $D=50\mu$m trap, with a cw He-Ne laser excitation. The emission
lines at a zero red shift are due to the direction exciton
emission. The bias is 4V and the excitation power is 30$\mu W$.
The dashed lines mark the trap boundaries. (c) The extracted
dipolar exciton density at the center of the excitation spot as a
function of the cw laser excitation power on the sample surface
for free expanding and trapped excitons. The applied bias is
6V.}\label{figure4}
\end{center}
\end{figure}

To show the dramatic effect of the exciton trapping, we now
compare free expanding dipolar excitons to trapped ones. When the
bias is applied uniformly to the whole sample (via a very large
gate, millimeters in size), the dipolar excitons that are
optically excited at a small spot diameter ($\sim$30 $\mu$m)
expand as a result of dipolar repulsion due to the radial density
gradient. For a dipolar exciton lifetime that exceeds 1 $\mu$s,
the exciton fluid can rapidly expand to hundreds of microns in
diameter \cite{VorosPRL2005,RapaportPRB2006}. This is shown in the
image of Fig.~\ref{figure4}a, which was taken by spectrally
resolving the emission from a slice passing through the excitation
spot created using a cw He-Ne laser. Here, the vertical axis shows
the spatial expansion of the dipolar exciton fluid and the
horizontal axis represents the PL energy, measured as red shift
relative to the direct exciton transition energy. As expected, at
the center of the excitation spot, the steady state density of
dipolar excitons is higher, thus the applied bias is partially
screened, leading to a slightly blue shifted emission compared
with positions further away from the excitation spot, where the
exciton density decreases and thus the emission corresponds to
that of dilute dipolar excitons. In contrast, the PL coming from a
50$\mu$m diameter Xtrap, \emph{under the same excitation
conditions} is very different, as is shown in Fig.~\ref{figure4}b.
It clearly shows no PL outside the trap, and a larger blue shift
(compared to the free excitons), indicating higher exciton
density.

\vspace*{0cm}

Neglecting small corrections from dipole-dipole interaction, the
dipolar exciton PL energy blue shift (compared to the dilute case)
is proportional to the fluid density due to the density dependent
screening of the external applied field by the dipoles
\cite{RapaportPRB2005,RapaportPRB2006}. This dependence can be
written as: $\Delta \varepsilon=(4\pi e^2z_0/\epsilon)n_X$, where
$n_X$ is the local exciton density and $\epsilon$ is the
background dielectric constant. This allows for an approximate
calibration of the experimental dipolar exciton density. In
Fig.~\ref{figure4}c, we present the steady-state exciton density
as a function of the excitation power at the sample surface for
various gate sizes, under the same experimental conditions. The
dipolar exciton density is thus obtained by monitoring the blue
shift, $\Delta \varepsilon$, of the emission from the center of
the PL profile as the excitation intensity increases.
Fig.~\ref{figure4}c shows that due to the driven expansion induced
by the dipolar repulsion, the achievable density of free dipolar
excitons (with no trap) is always far below that can be obtained
in the Xtraps. At high excitation intensities, exciton densities
as high as $1.8\times 10^{11}cm^{-2}$ are observed in Xtraps,
approaching the limit where the inter-exciton spacing becomes
comparable to the exciton Bohr radius. The increase of exciton
density in Xtraps becomes sublinear at high excitation
intensities. The mechanism for this sublinear behavior seems to be
a partial filling of the trap as the power is increased, and
consequently an expected decrease in the exciton lifetime
\cite{AlexandrouPRB1990} ($n_X=G\tau_X(G)$, where $G$ is the
exciton photogeneration rate and $\tau_X$ is the filling dependent
and hence the power dependent exciton lifetime). With the same
excitation powers, a larger steady exciton density is observed in
the smaller Xtraps, as optically excited excitons spread over a
smaller trap area. It is then clear that by making a high quality
exciton trap, one can indeed achieve a confined, high-density,
long-lived, and homogeneously distributed exciton fluid.

We now briefly discuss the expected unique physical properties of
the dipolar exciton gas that is confined in the Xtrap. The
De-Broglie temperature, which mark the approximate transition to
the regime where quantum statistics starts to play an important
role, is $T_{Xqs}\simeq 0.7(n_X/10^{10})K$ (where $n_X$ is in
units of cm$^{-2}$). On the other hand, the classical
nearest-neighbor dipole-dipole interaction term, calculated for a
square dipolar exciton lattice in equilibrium is
$\varepsilon_{dd}^{eq}/k_B=(4d_X^2/(k_B\epsilon))n_X^{3/2}\simeq
1.3(n_X/10^{10})^{3/2}K$, and the harmonic potential for small
vibrations is approximated to be
$\hbar\omega_X/k_B=\sqrt{18\hbar^2d_X^2/(k_B^2\epsilon
m_X)}n_X^{5/4}\simeq 1.57(n_X/10^{10})^{5/4}K$ \cite{parameters}.
As one can see, it turns out that at the relevant experimental
densities in the Xtrap ($n_X\sim 10^{10}$cm$^{-2}$), all these
physical quantities are of the same order of magnitude, which can
lead to some very interesting consequences in terms of the
possible thermodynamic phases of the excitons, such as a possible
competition of the quantum superfluid phase with the exciton
crystal phase. Note that crystallization of particles with a net
repulsive interaction can only happen within a confined volume and
thus is expected to be unique to the dipolar excitons in the trap.
What happens to a 2D dipolar exciton fluid in this regime is yet
to be explored.

%-----------------------------------------------------------------------------


\begin{thebibliography}{19}
\bibitem{sample_details} The two samples consist of a DQW
structure grown on an n$^+$ doped GaAs substrate. The two GaAs QWs
are 100\AA\ wide separated by a 40\AA\  Al$_{0.45}$Ga$_{0.55}$As
barrier and are cladded by 500\AA\  Al$_{0.45}$Ga$_{0.55}$As
barriers. The nominal separation between the two QW centers is
$z_0=140\AA$. In the absence of external bias, the direct exciton
transition is at 1.5567 eV (797 nm). The indirect dipolar exciton
transition energies qouted in this paper are relative to the above
value.
\bibitem{parameters} We use the following parameter values:
$m_X=0.2m_e$, $\epsilon=10$ and $z_0=14nm$.
\end{thebibliography}
\end{document}